\documentclass{article}

\usepackage[english]{babel}
\usepackage{braket}

\usepackage[letterpaper,top=2cm,bottom=2cm,left=3cm,right=3cm,marginparwidth=1.75cm]{geometry}
\usepackage{mathtools}

\usepackage{amssymb}
\usepackage{amsmath}
\DeclareSymbolFont{usualmathcal}{OMS}{cmsy}{m}{n}
\DeclareSymbolFontAlphabet{\mathcal}{usualmathcal}
\allowdisplaybreaks

\usepackage{graphicx}
\usepackage[colorlinks=true, allcolors=blue]{hyperref}

\usepackage{natbib}
\setcitestyle{numbers,square,comma}

\begin{document}

\title{Instabilities of the continuous superradiant laser}
\author{Bruno Laburthe-Tolra$^{1*}$, Martin Robert-de-Saint-Vincent$^{1}$, Benjamin Pasquiou$^{1}$}

\maketitle

\begin{center}
{ 1} Laboratoire de physique des lasers, CNRS, UMR 7538, Université Sorbonne Paris Nord, F-93430 Villetaneuse, France
\\

${}^\star$ {\small \sf bruno.laburthe-tolra@univ-paris13.fr}
\end{center}

\begin{abstract}
We investigate the intensity stability of the superradiant laser. Our study focuses on the architecture where a continuous beam of atoms in an electronically excited state crosses the mode of a high-finesse Fabry-Perot cavity, which has been proposed as a new architecture of an active optical clock. We show that such superradiant laser can become unstable and develop chaotic behavior. We derive an analytical criterion for this instability and find that it may only occur when the lifetime of photons in the cavity is significantly shorter than the lifetime of atoms. This criterion allows for refining the necessary parameters to run a superradiant laser as a frequency reference in the optical domain. In particular, we point-out the consequences of the instability on intensity fluctuations and laser linewidth. On the other hand, we also point out that the superradiant laser, when in the unstable regime, can become an interesting playground for studying chaos. At the mean-field level, there is a direct mapping to the Bénard instability associated with fluid turbulence; however quantum fluctuations associated with photon out-coupling and atom re-filling substantially modify the expected behaviors. Finally, we point-out the existence of a regular self-pulsing regime at large atom numbers.
\end{abstract}

\section{Introduction}

The prospect of a superradiant laser in the bad-cavity regime has attracted considerable attention both from the experimental point of view \cite{Norcia2016SuperradianceStrontiumClock,Pan2020CsActiveClockSuperradiant,Laske2019PulseDelaySuperradiantCa,Schaeffer2020LasingSrMOT3P1,Fama2024ccq} and the theoretical point of view \cite{Chen2005ActiveClockFirstProposal,Meiser2009MilliHzLaser,Liu2020RuggedSuperRadiantOven,Laburthe2023cal}. This is in part motivated by possible applications as an optical frequency standard, or to improve the stability of an ultra-stable laser (as it would truly realize a maser in the optical domain). The bad-cavity regime is useful for this prospect since the light emission is expected to be largely independent of mirror vibrations, and the laser frequency is mostly set by the atomic resonance frequency. However, it has long been known that lasers in the bad cavity regime can become unstable and display chaotic behavior \cite{Haken1975abh,Haken1986}. To our knowledge, this limitation has not yet been discussed by the community investigating superradiant lasers for metrological applications, although other types of instabilities, $e.g.$, due to atom motion \cite{Jaeger2021SuperRadiantThermalBeamCavity,Jager2021rab} or for separate ensembles of atoms \cite{Patra2019ddd}, have been predicted.

In this work, we show that the superradiant laser can indeed become unstable at large atom numbers, in a parameter regime relevant to the experiments being built. Our work focuses on a mean-field description, but we have verified that such instability holds in a semi-classical model based on the Truncated-Wigner approximation. In the first part of this paper, we study this instability numerically using Monte-Carlo simulations. We then develop an analytical model that reproduces the numerical results very well. A stability analysis of such a simplified model shows that the instability arises due to a dynamical instability. We provide analytical laws for the dynamical instability threshold. This allows us to explicitly define the experimental parameters that must be avoided to use the superradiant laser as a frequency reference, and we discuss the impact of the instability for both the continuous beam and the repumped superradiant laser architectures. We also point out the interest of the superradiant laser in the unstable regime, which could become a new system to study chaos and turbulent behavior, in the regime where turbulence will be significantly impacted by quantum fluctuations. We finally discuss a more stable regime that appears at very large atom numbers, in which case the turbulent laser gradually turns into a self-pulsing laser. This bears similarities with recently observed periodic superradiance using rare-earth ion-doped crystals \cite{Hara2024psi,Hara2026aan,Xie2026oot}. 

\section{The cw superradiant laser can become unstable - Monte-Carlo simulations}

\subsection{Effective Hamiltonian}

We begin the modeling of a superradiant laser as in our previous work \cite{Laburthe2023cal}, by considering $N$ atoms with ground and excited electronic levels $\ket{g}, \ket{e}$ resonantly coupled to a cavity. We introduce the resonant atom-cavity coupling Hamiltonian:
\begin{equation}
    H_0= g \sum_{j=1}^N (\ket{e_j} \bra{g_j} b  + h.c.),
\end{equation}
where $\ket{g_j}, \ket{e_j}$ are the states associated with each atom $j$, and $h.c.$ stands for the Hermitian conjugate. $g$ is the single-atom cavity-light interaction parameter, assumed to be identical for all atoms. $b$ is the destruction operator for a photon of the cavity mode. Summation over $j$ is made over all atoms present in the cavity mode. We define the spin operators $\ket{e_j} \bra{g_j} = s^{+}_j $, $\ket{g_j} \bra{e_j} = s^{-}_j$, and $\frac{1}{2} \left( \ket{e_j}\bra{e_j}-\ket{g_j}\bra{g_j}  \right) = s^{z}_j$. To describe the leaking of photons through the cavity mirrors, we follow the approach in Ref.~\cite{Laburthe2023cal} and find the effective Hamiltonian $H=H_0 +\frac{\kappa}{2 i} (b^{\dag} \langle b \rangle -b \langle b^{\dag} \rangle)$. Photons losses are thus described by a single leak-rate parameter $\kappa$ corresponding to the cavity photon lifetime. The symbols $\langle \rangle$ stands for the quantum expectation values. From this expression we compute the evolution of the system (using $i \frac{dA}{dt} = \left[ A,H \right]$). We make the mean-field approximation and thus neglect correlations between the atomic degrees of freedom and the cavity field, and we find:
\begin{eqnarray}
 i  \frac{d \langle b \rangle}{dt} &=& g \sum_j \langle s^-_j \rangle - \frac{i}{2}  \kappa  \langle b \rangle \nonumber \\
  i  \frac{d \langle b^{\dag} \rangle}{dt} &=& - g \sum_j \langle s^+_j \rangle - \frac{i}{2}  \kappa  \langle b^{\dag} \rangle \nonumber \\
 i  \frac{d \langle s^-_j \rangle}{dt} &=& -2 g  \langle b   \rangle \langle s^z_j \rangle \nonumber \\ 
 i  \frac{d \langle s^+_j \rangle}{dt} &=&  2 g  \langle b^{\dag} \rangle \langle s^z_j \rangle \nonumber \\ 
 i  \frac{d \langle s^z_j \rangle}{dt} &=&    g  \langle b   \rangle \langle s^+_j \rangle - g \langle b^{\dag} \rangle \langle s^-_j \rangle
 \label{elementarydynamics}
 \end{eqnarray}

\subsection{Monte-Carlo simulations}

We perform Monte-Carlo simulations of the superradiant laser dynamics for a given intracavity atom number $N$ by simulating the $3 N + 2$ equations of Eqs.~\ref{elementarydynamics}. In order to take into account atoms leaving and entering the cavity, we segment the simulation in a large number of temporal chunks, each of duration $\tau = 1 / \Gamma$, corresponding to the typical insertion time of an atom in the cavity. At the end of each segment, we randomly choose which atom $M$ leaves the cavity, and replace it by a ``fresh" atom for which initially $\langle s^z _M \rangle =1/2$ and $\langle s^x_M \rangle= \langle s^y_M\rangle=0$. We then continue the evolution on the next time-segment. We label this random choice method as ``method 1" in the results of Fig.~\ref{agreement}. 

In this procedure, the probability for one atom to leave the cavity after a time exactly $p \, \tau$, $p \in \mathbf N$, is $P(p \, \tau) \approx \frac{1}{N} \left( 1-\frac{1}{N} \right)^p \approx \frac{1}{N} e^{-\frac{p}{N}}$. Thus, the $1/e$ lifetime of an atom in the cavity is $N \tau$, which defines the transit rate $\Gamma_R = \Gamma / N$. This quantity $\Gamma_R \propto \bar{v} / w$ is one of the key elements characterizing the experiments, with $\bar{v}$ qualitatively representing the atoms' average velocity and $w$ the transverse size of the cavity. Note that, while in an experiment the mode of the cavity will typically be Gaussian, our modeling approximates the mode field by a Heaviside function , $i.e.$, atoms are supposed to be identically coupled to the cavity over a physical distance $w$ perpendicular to the cavity axis. Therefore, the comparison between our model and an experiment will necessarily be qualitative.

We point out the following additional difficulty concerning the velocity distribution that arises when using this Monte-Carlo sampling. Randomly choosing which atoms leaves the cavity every $\tau$ corresponds to an atom beam with velocity probability distribution  $P(v) = \frac{w  \Gamma}{N v ^2} e^{ -\frac{w \Gamma}{N v} } $ for $v< \Gamma w$ and $P(v)=0$ for  $v> \Gamma w$, from which we deduce the average velocity $\bar{v}=\Gamma_R w \times \Gamma_E \left( 0,\frac{1}{N} \right)$, where $\Gamma_E$ is the Euler Gamma function. The Euler Gamma function contains a logarithmic divergence as a function of the atom number, so that the average velocity is an ill-defined quantity and diverges for $N \rightarrow \infty$. This is due to the fact that the random choice allows for the unlikely possibility for atoms to have a velocity larger than $N \times \Gamma_R w$ with a non-negligible ($\approx \frac{1}{N}$) probability. While this difficulty does not greatly modify the average velocity for realistic atom numbers, we have also explored another Monte-Carlo sampling that does not suffer from the logarithmic divergence and gives very similar results. 

The second Monte-Carlo sampling corresponds to a different random choice for atom losses. In this approach $\frac{1}{10} ^{th}$ of the atoms do not undergo losses at the beginning of their cruise through the cavity mode. Then, the ``older" $\frac{9}{10}^{th}$ of the atoms undergo random losses with equal probability. This corresponds to a velocity distribution $P(v) = \frac{10 w  \Gamma}{9 N v ^2} e^{ -\left(\frac{w \Gamma}{ v} -\frac{N}{10} \right) \frac{10}{9N} } $ for $v< 10 w \Gamma_R$ and $0$ otherwise, and an average velocity $\bar{v} = \frac{10}{9}  e^{\frac{1}{9}}   \Gamma_R   w \Gamma_E \left(0,\frac{1}{9} \right) \approx 2 \Gamma_R w$; the velocity standard deviation is comparable to the average velocity (which conveniently describes the typical spread of velocities from an oven, a 2D magneto-optical trap, or out of a Zeeman slower apparatus.) This small change in the distribution of atoms losses and its corresponding Monte-Carlo sampling solves the logarithmic divergence, while it does not drastically change the results of the simulation. The result in Fig.~\ref{agreement} shows the comparison between this second random choice ``method 2" and the first.

\subsection{Results} 

\begin{figure}
\centering
\includegraphics[width= 0.8\columnwidth]{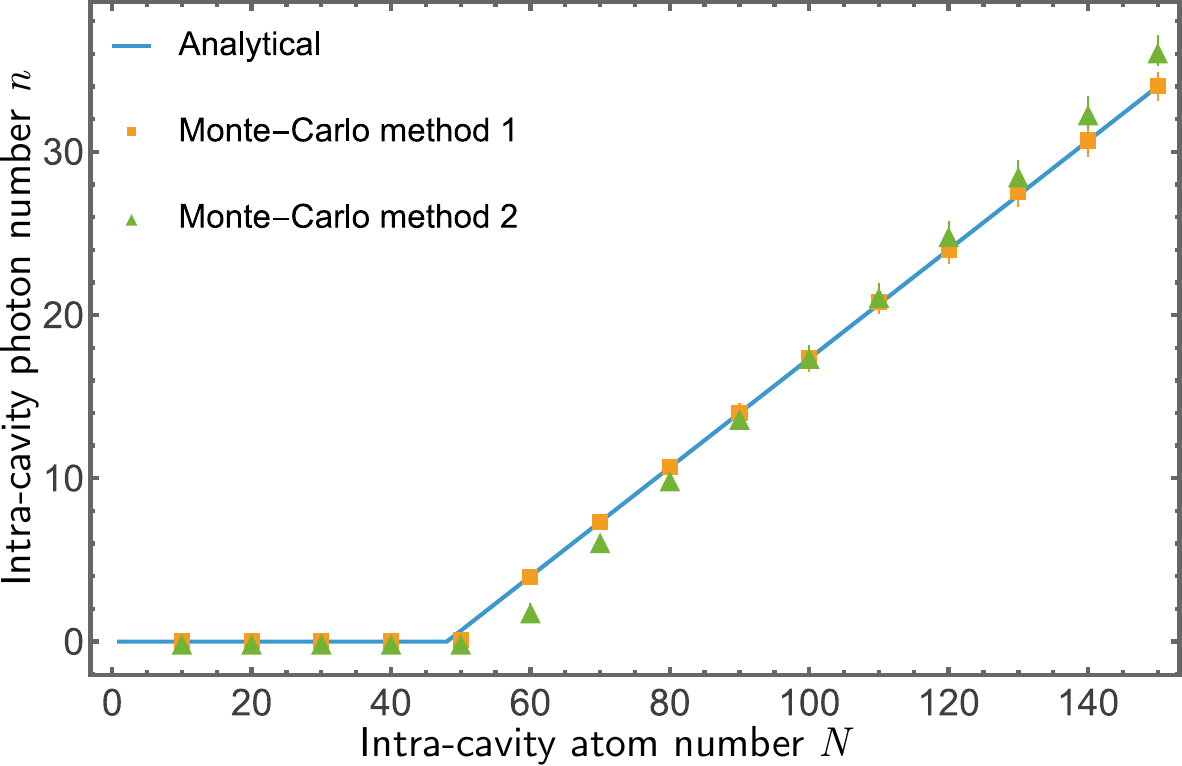}
\caption{\setlength{\baselineskip}{6pt} {\protect\scriptsize Steady-state superradiant laser power as a function of atom number $N$ inside the cavity mode. Yellow squares are the results of Monte-Carlo simulations of Eqs.~\ref{elementarydynamics} using the random choice method 1, see main text, while green triangles are the results using method 2. The blue line is the analytical formula from Ref.~\cite{Laburthe2023cal} (eq. 29-30). The parameters for all these results are $g/2\pi = 1/2$, $\kappa/2\pi = 6$, and $\Gamma_R/2\pi = 4$. The cavity photon number is calculated by averaging the last 20 percent of the simulated dynamics, running for a total time $60 / \Gamma_R$. For each point, the error bars show the standard deviation obtained from 140 Monte-Carlo realizations.}} 
\label{agreement}
\end{figure}

First, as a benchmark of our model and Monte-Carlo method, we explore the steady-state power of the superradiant laser as a function of atom number, shown in Fig.~\ref{agreement}. We observe the typical lasing threshold, in very good agreement with theoretical predictions \cite{Liu2020RuggedSuperRadiantOven, Laburthe2023cal}, giving lasing for atom numbers $N > N_c \equiv \frac{\Gamma_R \kappa}{2 g^2}$. 

\begin{figure}
\centering
\includegraphics[width= 0.8\columnwidth]{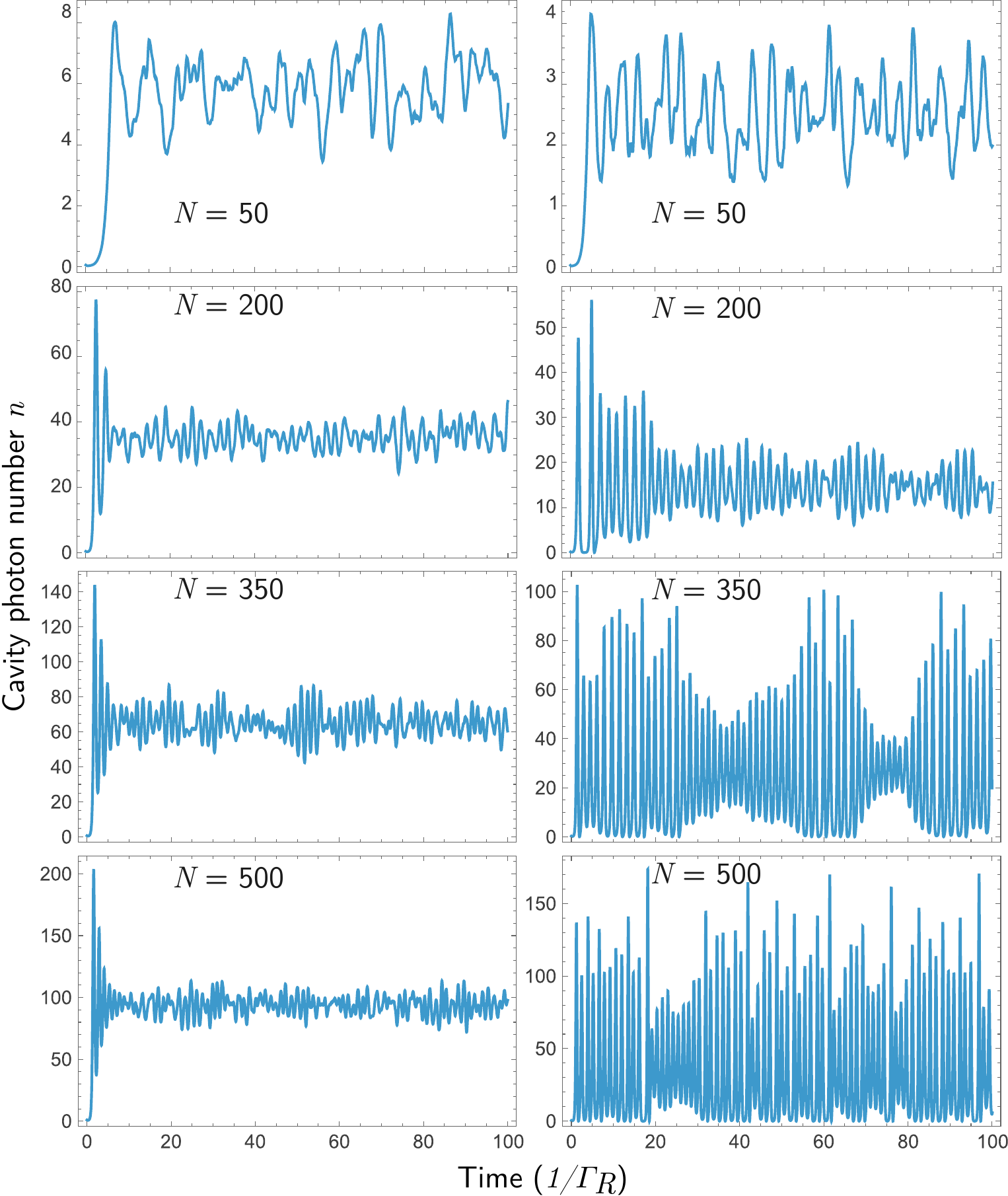}
\caption{\setlength{\baselineskip}{6pt} {\protect\scriptsize Examples of superradiant laser dynamics from Monte Carlo simulations of Eq.~\ref{elementarydynamics}. Left column: $g/2\pi = 1$, $\kappa/2\pi =10$, $\Gamma_R/2\pi =3.9$; right column: $g/2\pi = 0.65$, $\kappa/2\pi =10$, $\Gamma_R/2\pi =1.65$. Both cases have the same critical atom number for lasing ($N_c = \Gamma_R \kappa / 2 g^2 \approx 20$). The left column correspond to the stable case ($\kappa < 4 \Gamma_R$) and the right column to the unstable case ($\kappa > 4 \Gamma_R$). In the latter case, the critical atom number $N_c^*$ for instability, see Eq.~\ref{criterioninstability}, is $N_c^* = \frac{\kappa^2 (8 \Gamma_R + \kappa)}{4  g^2  (\kappa-4 \Gamma_R )} \approx 400$. Each solid line is the result of a Monte-Carlo simulation (see paragraph).}} \label{timedependentmontecarlo}
\end{figure}

We then study the laser dynamics in two cases, see Fig.~\ref{timedependentmontecarlo}: both cases feature the same critical atom number $N_c$ for the lasing threshold; they differ in the parameters $g^2$ and $\Gamma_R$, which are scaled by a parameter $c$ that keeps $\Gamma_R / g^2 \propto N_c$ constant. As we shall now see, although the cavity parameters are not vastly different, the behavior of the laser as a function of atom number strongly differ. The time-dependent intra-cavity photon number is plotted as a function of time for various intra-cavity atom number $N$. For each panel, the simulation starts with an empty cavity. In both cases, for a small number of atoms above the lasing threshold criterion, we observe relaxation oscillations that rapidly converge toward a steady state. For large numbers of atoms, we observe a very different behavior between the two cases. In the case with the highest $\Gamma_R$, the laser remains rather stable, while in the other case, the laser develops intensity oscillations that become more and more pronounced for increasing atom number, until the laser becomes basically pulsed. This observation shows that the intensity stability of the superradiant laser can be strongly sensitive to experimental parameters. 

\subsection{Stability diagram}

\begin{figure}
\centering
\includegraphics[width= 0.65\columnwidth]{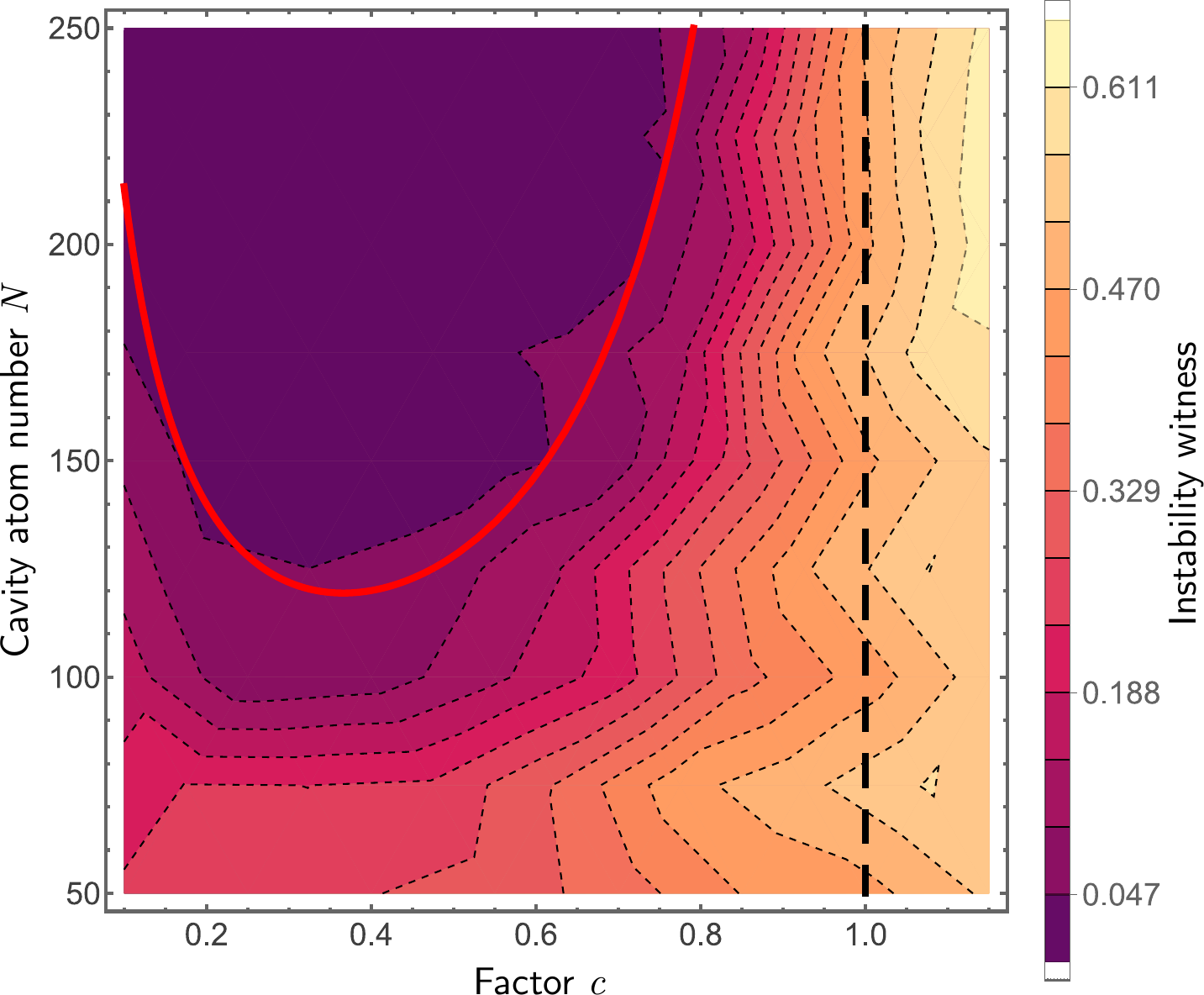}
\caption{\setlength{\baselineskip}{6pt} {\protect\scriptsize We plot the instability witness (see main text) estimated from Monte Carlo simulations of Eqs.~\ref{elementarydynamics}, as a function of the atom number and the instability parametrization factor $c$, with $g/2\pi = \sqrt{c} /2$, $\kappa/2\pi =4$, $\Gamma_R/2\pi =c$. The parametrization is chosen so that the atom number threshold for lasing ($N_c = \Gamma_R \kappa  /2 g^2 \approx 6$) is independent of $c$. The black dashed line shows the limit $ 4 \Gamma_R = \kappa$ above which the laser is always stable, irrespective of atom number (right side). The red line depicts the critical atom number $N_c^*$ for the instability to develop as a function of $c$, $N_c^* = \frac{\kappa^2 (8 \Gamma_R + \kappa)}{4  g^2  (\kappa-4 \Gamma_R )}$, see Eq.~\ref{criterioninstability}.}} 
\label{instabilitydiag}
\end{figure}

We now explore the stability diagram of the superradiant laser, by systematically performing time-dynamics simulations as in Fig.~\ref{timedependentmontecarlo} for varying $N$ and scaling parameter $c$. As previously, we scale $g^2$ and $\Gamma_R$ using $c$, so that the critical number $N_c$ is constant. The results are shown in Fig.~\ref{instabilitydiag}, where we plot a witness of instability. This witness consists in the minimum observed photon number during the last $10\%$ of the simulated dynamics (running up to times of $60/\Gamma_R$) divided by the average photon number, and therefore characterizes intensity stability in a simple fashion. This ratio approaches zero when the instability is reached. Figure~\ref{instabilitydiag} shows that the superradiant laser remains stable for any number of atoms for large enough $c$, but an instability develops for large $N$ and small $c$. 

\section{Interpretation}

\subsection{Effective model}

\subsubsection{Basic equations}

To interpret the observations above, we develop a simplified model, building on the one of Eq.~\ref{elementarydynamics}. This time we explicitly write in the effective model the insertions and losses of atoms into and out of the cavity, using the parameters $\Gamma$ and $\Gamma_R$ that we have already introduced. Again, we assume that all atoms are identically coupled to the cavity mode. We neglect any velocity along the cavity axis. In practice, the key point is that the atoms must move by less than half the wavelength of the light along the cavity axis as they cross its mode, which ensures to remain far from the instabilities described in \cite{Jaeger2021SuperRadiantThermalBeamCavity,Jager2021rab}. We also neglect the spontaneous emission rate $\gamma_s$, which is a good approximation when $\Gamma_R,\,\kappa \gg \gamma_s $.

We continue to make the mean-field approximation and define the sums of expectation values $S^{+,-,z} = \sum_j \langle s^{+,-,z}_j \rangle$. For concision, in the remainder of this article, we rewrite $\langle b \rangle \rightarrow b$, and we do not write the equations for $\frac{db^{\dag}}{dt}$ and $\frac{dS^+}{dt}$. Following the results of Eq.~\ref{elementarydynamics} and Ref.~\cite{Laburthe2023cal}, we obtain:
\begin{eqnarray}
    i  \frac{db}{dt} &=& g S^-  -\frac{i}{2}  \kappa  b \nonumber \\
   i  \frac{dS^-}{dt} &=&  -2 g  b S^z - i \Gamma_R S^- \nonumber \\
 i  \frac{dS^z}{dt} &=&  g b S^+ - g  b^{\dag}  S^- - i \Gamma_R S^z+ \frac{i}{2} \Gamma ,
 \label{averagedynamics}
 \end{eqnarray}
where we recall that $\Gamma = \Gamma_R N$.

\subsubsection{Three types of lasers}

From Eqs.~\ref{averagedynamics} it is easy to define three well-known families of lasers, depending on whether the coherences are set by the field (standard laser) \cite{Haken1986}, the atoms (superradiant laser) \cite{Chen2005ActiveClockFirstProposal, Meiser2009MilliHzLaser}, or whether the system is in the cross-over regime \cite{Tieri2017CrossoverLasingSuperradiance, Debnath2018SuperradiantNarrowLinewidth}. As we shall now see, our minimalist model covers all three regimes and allows us to precisely pinpoint which basic approximation is made in each regime. 

\begin{itemize}
    \item The first regime is when $\kappa \ll \Gamma_R$ (\textit{good-cavity regime}). We perform adiabatic elimination of the atomic coherences setting $\frac{dS^-}{dt}=\frac{dS^+}{dt}=0$. We choose the gauge where b is real (which is possible in the mean-field description), and define $n = b^2$ the photon number and $D = 2 S^z$ the population inversion, and find:
\begin{eqnarray}
\frac{dn}{dt} &=&  W n D - \kappa n \nonumber \\
\frac{dD}{dt} &=&  - 2 W n D -\Gamma_R D + \Gamma, 
\label{rateequation}
\end{eqnarray} which are the well-known rate equations for the single-mode laser, with pumping rate $W=\frac{2 g^2}{\Gamma_R}$. Thus, the usual rate equations for a laser fundamentally rely on the adiabatic elimination of the atomic coherences. In the multimode regime, deriving the rate equations from first principle also implies neglecting the phase relations between the laser modes \cite{Haken1986}.
\item  The second regime is when $\kappa \gg \Gamma_R$ (\textit{bad-cavity regime}), in which case it is natural to perform instead the adiabatic elimination of the electric field, $\frac{db}{dt}=\frac{db^{\dag}}{dt}=0$. Adiabatic elimination of the field leads to the basic superradiant laser equations \cite{Laburthe2023cal}:
\begin{eqnarray}
\frac{dS^-}{dt} &=& \frac{4g^2}{\kappa} S^- S^z - \Gamma_R S^- \nonumber \\
\frac{dS^z}{dt} &=&  - \frac{4g^2}{\kappa} S^- S^+  - \Gamma_R S^z + \frac{\Gamma}{2}
\label{SRequation}
\end{eqnarray}
\item The intermediate, \textit{cross-over regime} is the one we are interested in. For the two previous regimes, since equations were defined in a real plane (two real variables) and cannot reach infinite values, the Poincaré-Bendixson theorem implies that no chaos can occur. As we shall see below, this is not the case when no adiabatic elimination is performed; the third, intermediate regime includes both regular and chaotic behaviors. Note that Ref.~\cite{Jager2020sop} obtains chaotic behavior even in the superradiant regime. However, this phenomenon appears when including the physics associated with atomic recoil, which introduces an additional degree of freedom.
\end{itemize}

\subsection{Dynamical instability}

\subsubsection{Instability threshold}

Coming back to the general expression of Eqs.~\ref{averagedynamics}, and choosing the gauge where $b$ is real and $S^-$ imaginary, the steady-state solutions are (for $N>N_c$):
\begin{equation}
    \left\{ b_0 = \pm \frac{\sqrt{2 g^2 \Gamma_RN-\Gamma_R^2 \kappa}}{2 g \sqrt{\kappa}}, \,  S_0^- = \pm i\frac{\sqrt{\kappa} \sqrt{2 g^2 \Gamma_RN-\Gamma_R^2 \kappa}}{4 g^2}, \,  S_0^z = \frac{\Gamma_R \kappa}{4 g^2}\right\}
    \label{solstat}
\end{equation}
To investigate the stability of the system, we consider small oscillations around one of these steady states, $i.e.$ we write $b = b_0+\delta b$, $S^- = S_0^-+\delta S^-$, and $S^z = S_0^z+\delta S^z$, and linearize the resulting equations. We obtain $\left\{ \dot{\delta b}, \, \dot{\delta S^-}, \, \dot{\delta S^z} \right\} = M \cdot \left\{\delta b, \, \delta S^-, \, \delta S^z\right\}$, with 
\begin{equation}
  M=  \left(
\begin{array}{ccc}
 -\frac{\kappa}{2} & - g & 0 \\
 -\frac{\Gamma_R \kappa}{2 g} & -\Gamma_R & -\frac{\sqrt{\Gamma_R \left(2 g^2 N -\Gamma_R \kappa\right)}}{\sqrt{\kappa}} \\
 -\frac{\sqrt{\kappa} \sqrt{\Gamma_R \left(2 g^2 N-\Gamma_R \kappa\right)}}{2 g} & \frac{\sqrt{\Gamma_R \left(2 g^2 N-\Gamma_R \kappa\right)}}{\sqrt{\kappa}} & -\Gamma_R \\
\end{array}
\right)
\end{equation}

An instability occurs when one eigenvalue of this matrix has a positive real part. (This linear stability criterion is motivated by the Hartman–Grobman theorem.) We find that the criterion for the instability, corresponding to the real part of the eigenvalue becoming positive, is 
\begin{eqnarray}
    \kappa &>& 4 \Gamma_R \nonumber \\ 
    N& >&\frac{\kappa^2 (8 \Gamma_R + \kappa)}{4  g^2  (\kappa-4 \Gamma_R )} \equiv N_c^*.
    \label{criterioninstability}
\end{eqnarray}
The analytical threshold $N_c^*$ for instability is in good agreement with our Monte-Carlo results of Fig.~\ref{instabilitydiag}. When $\kappa < 4 \Gamma_R$, the laser is always stable.

In order to further characterize the stability of the laser, we introduce:
\begin{eqnarray}
A^2/g^6 &=&\Gamma_R (-\Gamma_R (8 \Gamma_R - 3 \kappa)^2 \kappa^4 (8 \Gamma_R + \kappa) + 
       8 g^2 \kappa^3 (256 \Gamma_R^3 - 152 \Gamma_R^2 \kappa + 5 \Gamma_R \kappa^2 + 2 \kappa^3) N \nonumber \\ 
       & & - 16 g^4 \Gamma_R \kappa (16 \Gamma_R^2 + 128 \Gamma_R \kappa - 71 \kappa^2) N^2 + 
       512 g^6 \Gamma_R^2 N^3) \nonumber \\
B &=& g^3 \sqrt{\kappa} (-\kappa (-8 \Gamma_R + \kappa) (-8 \Gamma_R^2 + 11 \Gamma_R \kappa + \kappa^2) + 
     36 g^2 \Gamma_R (4 \Gamma_R - 5 \kappa) N) + 3 \sqrt{3} A 
\label{eq_A_and_B}
\end{eqnarray}
The three eigenvalues of $M$ read:
\begin{eqnarray}
   \gamma_1 &=& \frac{g \left(2 \Gamma_R \left(\kappa^2-12 g^2 N\right)+16 \Gamma_R^2 \kappa+\kappa^3\right)}{6 \sqrt[3]{B} \sqrt{\kappa}}+\frac{\sqrt[3]{B}}{6 g \sqrt{\kappa}}+\frac{1}{6} (-4 \Gamma_R-\kappa) \nonumber \\
   \gamma_2 &=&    \frac{\left(-1- i \sqrt{3}\right) g \left(2 \Gamma_R \left(\kappa^2-12 g^2 N\right)+16 \Gamma_R^2 \kappa+\kappa^3\right)}{12 \sqrt[3]{B} \sqrt{\kappa}}+\frac{ \left(-1+i\sqrt{3}\right) \sqrt[3]{B}}{12 g
   \sqrt{\kappa}}+\frac{1}{6} (-4 \Gamma_R-\kappa) \nonumber \\
 \gamma_3 &=&\frac{ \left(-1+i\sqrt{3}\right) g \left(2 \Gamma_R \left(\kappa^2-12 g^2 N\right)+16 \Gamma_R^2 \kappa+\kappa^3\right)}{12 \sqrt[3]{B} \sqrt{\kappa}}+\frac{\left(-1- i \sqrt{3}\right) \sqrt[3]{B}}{12 g
   \sqrt{\kappa}}+\frac{1}{6} (-4 \Gamma_R-\kappa) .
\end{eqnarray}
Two of these eigenvalues are complex conjugate (exactly which ones depends on the sign of $B$). The other eigenvalue is real, and is always negative. Therefore there are two excitation modes, and two real values define their damping constants.

\subsubsection{Rate of instability}

We point out that $B>0$ for $N>\frac{\kappa (16 \Gamma_R^2 + 2 \Gamma_R \kappa + \kappa^2)}{24 g^2 \Gamma_R}$. Using this, we find that the growth rate of the instability, given by the real part of the complex-conjugate eigenvalues, reads at second order in $1/N$ when $N \rightarrow \infty$: 
\begin{equation}
\gamma_i \approx \frac{1}{4}  (-4 \Gamma_R + \kappa) - \frac{\kappa^2 (2 \Gamma_R^2 - 3 \Gamma_R \kappa + \kappa^2)}{8 g^2 \Gamma_R N } -\frac{(7 \Gamma_R - 4 \kappa) (\Gamma_R - \kappa) (2 \Gamma_R - \kappa) \kappa^4}{32 g^4 \Gamma_R^2 N^2} .
    \end{equation}

\subsubsection{Natural frequency of oscillation}
    
The oscillation frequency, given by the imaginary part of the corresponding eigenvalue, reads when $N \rightarrow \infty$:
\begin{equation}
   \gamma_{osc} \approx  g \sqrt{ \frac{2\Gamma_R}{\kappa} N} .
\end{equation}
Therefore, for atom numbers below the instability threshold $N_c^*$, the steady state of the superradiant laser is single-mode and continuous-wave, and any deviation from the steady state will lead to damped oscillations at the frequency $\gamma_{osc}$ with a damping rate $\gamma_i$. However, above the instability threshold any fluctuation will be exponentially amplified, which marks the onset of chaotic behavior.

\subsubsection{Physical interpretation}

We point out that for $\kappa \gg \Gamma_R$, $N>N_c^*$ corresponds to $N > \frac{\kappa^2}{4 g^2}$ and therefore to the breakdown of adiabatic elimination: $N \frac{g^2}{\kappa} > \kappa$ means that collective emission of atoms into the cavity is faster than the cavity leak rate, and indicates that the atoms have enough time to emit and reabsorb light before the light leaks out from the cavity. This provides a physical interpretation for the appearance of pulsed behavior.

\subsection{Connection to turbulence and to the Lorenz-Haken model}

\subsubsection{Mapping laser equations to the Lorenz model for turbulence}

We rewrite Eq.~\ref{averagedynamics} using variables renormalized to the steady-state values given by Eq.~\ref{solstat}, $\left\{ \beta , \Sigma^- , \Sigma^z \right\} \equiv \left\{ \frac{b}{b_0},  \frac{S^-}{S^-_0}, \frac{S^z}{S^z_0} \right\}$, and using the critical number for the lasing threshold $N_c = \frac{\Gamma_R \kappa}{2 g^2}$:
\begin{eqnarray}
    \frac{d\beta}{dt} &=&  \frac{\kappa}{2} \left( \Sigma^- - \beta \right) \nonumber \\
    \frac{d\Sigma^-}{dt} &=&  \Gamma_R \left( \beta \Sigma^z - \Sigma^- \right) \nonumber \\
    \frac{d\Sigma^z}{dt} &=& - \Gamma_R \beta \Sigma^- \left( \frac{N}{N_c} -1 \right) - \Gamma_R \left( \Sigma^z - \frac{N}{N_c} \right) .
     \label{averagedynamicsrenorm}
\end{eqnarray}
We point out the similarity of Eqs.~\ref{averagedynamicsrenorm} to the Lorenz-Haken model leading to instabilities and chaos in a traveling-wave laser \cite{Haken1975abh,Haken1986}:
\begin{eqnarray}
    \frac{dE}{dt} &=& \frac{\kappa}{2} \left(P - E \right) \nonumber \\
    \frac{dP}{dt} &=& \gamma_{\perp} \left(E D - P \right) \nonumber \\
    \frac{d D}{dt} &=& \gamma \left(- \Lambda EP - D  + \Lambda + 1 \right) ,
    \label{LHmodel}
    \end{eqnarray}
where $\kappa$ is the photon dissipation rate, $\gamma_{\perp}$ the polarization relaxation rate, and $\gamma$ the population difference relaxation rate. $E$ stands for the field, $P$ the polarization, and $D$ the population inversion. These equations are taken at resonance. Setting $P= \Sigma^-$, $D= \Sigma^z$, $E= \beta$, and for the specific case where  $\gamma=\gamma_{\perp}= \Gamma_R$, together with $\Lambda = \frac{N}{N_c} - 1$, there is an exact mapping between Eqs.~\ref{averagedynamicsrenorm} and Eqs.~\ref{LHmodel}. 

This mapping allows us to connect our results to the ones of the Lorenz-Haken model. This model indeed contains a dynamical instability and a transition toward a chaotic region when $\frac{\kappa}{2}> \gamma + \gamma_{\perp}$ (equivalent to $\kappa > 4 \Gamma_R$) at large $\Lambda$ (corresponding to large $N$). Interestingly, this constitutes a mapping between chaotic laser behavior and the Lorenz model for turbulence. 

On the other hand, in the opposite regime $\frac{\kappa}{2} < \gamma + \gamma_{\perp}$ the system is expected to remain stable. Note that the predicted transition to self-pulsing toward a mode-locked ultra-short pulses regime \cite{Haken1986} cannot be observed in our model that does not take into account the spatial dependence of the electric field along the cavity axis.

\subsubsection{Generalization to the repumped superradiant laser}

The connection between superradiance and turbulence is not specific to the case of the continuous atomic beam approach. Here, we consider the case of the superradiant laser in presence of a laser that incoherently pumps ground state atoms into the excited state, in which case each atom can emit more than one photon. Mean-field equations can be found in  Refs.~\cite{Tieri2017CrossoverLasingSuperradiance, Jager2017sto, Kirton2017sar,Dubey2024PhDThesis}:
\begin{eqnarray}
    i  \frac{db}{dt} &=& g S^-  -\frac{i}{2}  \kappa  b \nonumber \\
   i  \frac{dS^-}{dt} &=&  -2 g  b S^z - i \frac{\Gamma_1}{2} S^- \nonumber \\
 i  \frac{dS^z}{dt} &=&  g b S^+ - g  b^{\dag}  S^- - i \Gamma_2 (S^z - \frac{N}{2} d_0) ,
\end{eqnarray}
with $\Gamma_1 = R + \gamma_s + \nu$, $\Gamma_2 = R + \gamma_s$, and $d_0=\frac{R-\gamma_s}{R+\gamma_s}$, where $R$ is the repumping rate, $\gamma_s$ the spontaneous emission rate, and $\nu$ the rate of some additional dephasing of the atomic dipole.

We choose the gauge where $b$ is real and $S^-$ imaginary, and we define $\Gamma_3 = R-\gamma_s$. The steady-state solutions are:
\begin{equation}
    \left\{ b \to \pm \frac{\sqrt{4 g^2 \Gamma_3 N - \Gamma_1 \Gamma_2 \kappa}}{2 \sqrt{2 \kappa} g }, \, S^- \to \pm  \frac{i \sqrt{\kappa} \sqrt{4 g^2 \Gamma_3 N - \Gamma_1 \Gamma_2 \kappa}}{4 \sqrt{2} g^2}, \, S^z\to \frac{\Gamma_1 \kappa}{8 g^2} \right\} .
\end{equation}
Scaling the dynamical equations to those steady-state values and introducing $N_c^{r} = \frac{\Gamma_1 \Gamma_2 \kappa}{4 g^2 \Gamma_3}$, we find the renormalized equations:
\begin{eqnarray}
    \frac{d\beta}{dt} &=& \frac{\kappa}{2} \left( \Sigma^- - \beta \right) \nonumber \\
    \frac{d\Sigma^-}{dt} &=& \frac{\Gamma_1}{2} \left( \beta \Sigma^z - \Sigma^- \right) \nonumber \\
    \frac{d\Sigma^z}{dt} &=& - \Gamma_2 \beta \Sigma^- \left( \frac{N}{N_c^{r}}-1\right)  - \Gamma_2 \left(\Sigma^z - \frac{N}{N_c^{r}} \right).
\end{eqnarray}
These equations are of the same form as those shown previously. Therefore, the instabilities and the close connection to turbulence are also expected in the repumped case.

\section{The regimes of the continuous-beam superradiant laser}

This section describes the lasing regimes contained in our minimalist model. Those can be summarized by the following table:
\begin{center}
\begin{tabular}{ |c|c|c| } 
 \hline
  & $\kappa < 4 \Gamma_R$ & $\kappa > 4 \Gamma_R$ \\ 
  \hline
 $N<N_c$ & No laser light & No laser light \\ 
 $N > N_c$ and $N_c^* < 0$ & Stable laser & --- \\
 $N_c < N < N_c^*$ & --- & Stable laser \\
 $N>N_c^*$ & Stable laser  & Turbulent regime \\
 $\sqrt{\frac{N}{N_c}} \frac{\Gamma_R}{\kappa} \gg1$ &Stable laser & Regular pulses \\
 \hline
\end{tabular}
\end{center}

\subsection{Stable regime}

\subsubsection{Intensity fluctuations}

We consider the stable regime and investigate the intensity fluctuations as a function of the experimental parameters. To do so, we focus on the situation where $\kappa > \Gamma_R$ (the superradiant regime, in which the optical field adapts more or less rapidly to the atomic variables), and the laser is still stable ($i.e.$, $N<N_c^*$ when this criterion applies). We then resort to the following heuristic approach to provide an analytical estimate for light-intensity fluctuations. Specifically, we consider small deviations from the steady state that arise from the out-coupling of atoms from the cavity: we have $b=\frac{-2gi}{\kappa} S^-$ in the steady state, but the loss of individual atoms introduces fluctuations in $S^-$ by typically $1/2$ at a rate $\Gamma$. This random process introduces a random walk that leads to a drift of $<b>$. Taking into account only this process, the variance $\text{Var}(b)$ (characterizing the typical fluctuations of $b$ over a duration $t$ for a given experimental realization) would increase linearly as a function of time, \textit{i.e.}, $\frac{d\text{Var}(b)}{dt} \rvert_{loss} \propto \frac{g^2}{\kappa^2} \Gamma$. 

As we shall now see, the fluctuations of $b$ are damped out over a characteristic time scale that is set by the real part of the eigenvalues of $M$. When $\kappa \gg g$, the  above described fluctuations correspond to the excitation of a mode where $\delta b \ll \delta S^-$. Such mode corresponds to one of the eigenvalues of $M$, and in practice, for $\kappa \gg g$, the inverse timescale to restore equilibrium is set by the real part of the complex conjugate eigenvalues of $M$. Alternatively, when $\kappa \gtrsim g$, the fluctuations excite a mode for which  $\delta b \lesssim \delta S^-$, and the damping is then characterized by the inverse timescale associated with the only real eigenvalue of $M$. Therefore, in both cases, we postulate that $\frac{d\text{Var}(b)}{dt} \approx \frac{g^2}{\kappa^2} \Gamma - 2 \gamma_M  \text{Var}(b)$, where $\gamma_M =\lvert \Re(\gamma) \rvert$ is the real part of the corresponding eigenvalue $\gamma$, and the factor 2 is because the equation is on $\text{Var}(b)$ rather than on $b$.

\begin{figure}
\centering
\includegraphics[width= 0.95\columnwidth]{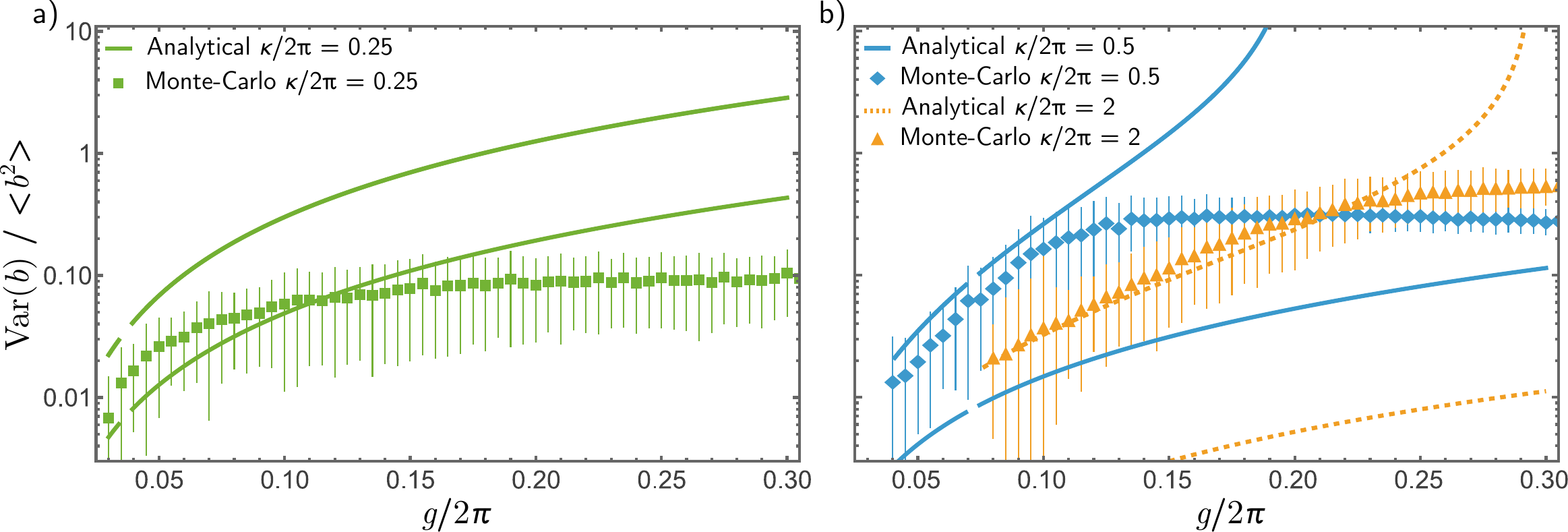}
\caption{\setlength{\baselineskip}{6pt} {\protect\scriptsize Superradiant laser intensity fluctuations in the stable regime. We plot the relative intensity noise $\text{Var}(b) / \langle b \rangle^2$ as a function of $g$ for three values of $\kappa$ and for $\Gamma_R/2\pi = 0.1$, $N=20$. a) In the regime with $\Gamma_R < \kappa < 4 \Gamma_R$, shown for $\kappa/2\pi = 0.25$, the laser is stable and the intensity fluctuations observed in the Monte-Carlo simulations of Eqs.\ref{elementarydynamics} (green squares) roughly follow the ones arising from one of the eigenvalues $\gamma_i$ (green straight lines) given by our simplified model (see Eq.~\ref{intensityfluctuation}). b) In the regime with $\kappa > 4 \Gamma_R$, we plot the Monte-Carlo simulation results for the cases $\kappa/2\pi = 0.5$ (blue diamonds) and $\kappa/2\pi = 2$ (orange triangles) along with their analytical counterparts (straight blue lines and dashed orange lines, respectively). We find good agreement between the observed fluctuations and the simplified model, until the relative intensity fluctuation approaches 1, \textit{i.e.}, when approaching the instability region (toward high $g$). Note that for both figures and at all values of $g$, there are always two eigenvalues with the same real part, so their associated curves overlap. The small gaps in the solid lines at $g / 2 \pi \approx 0.37$ (a) and $g / 2 \pi \approx 0.72$ (b) are due to jumps in the values of $\gamma_1, \, \gamma_2, \, \gamma_3$ when $B$ (of Eq.~\ref{eq_A_and_B}) changes sign. Error bars show the standard deviation for each point, obtained by averaging 280 Monte-Carlo simulations.}} 
\label{laserfluctuation}
\end{figure}

From this equation we deduce that the intensity fluctuations of the laser can be characterized by
\begin{equation}
    \frac{\text{Var}(b)}{\langle b \rangle^2} \approx \frac{ g^2}{\kappa \gamma_M}.
    \label{intensityfluctuation}
\end{equation}
We point out that $\frac{\text{Var}(b)}{\langle b \rangle^2} \approx \frac{1}{4} \frac{\text{Var}(n)}{\langle n \rangle ^2}$, assuming that $b$ follows a sufficiently narrow distribution of real values. Figure~\ref{laserfluctuation} represents the relative intensity fluctuation as a function of $g$ for several values of $\kappa$, and keeping $\Gamma_R < \kappa$. The solid lines are the results of Eq.~\ref{intensityfluctuation}, scaled by an overall vertical factor $x=0.3$, while the points represent the results of Monte-Carlos simulations. Each data point corresponds to an average over the results of 280 Monte-Carlo simulations. The panel a) is for $\kappa \approx g$ and the panel b) for $\kappa \gg g$. Both cases correspond to the excitation of two different modes, and the corresponding eigenvalues are used for the analytical formula. We generally observe a good agreement between the Monte-Carlo results and the formula given by Eq.~\ref{intensityfluctuation}.

The expression Eq.~\ref{intensityfluctuation} is similar to the result for standard lasers $\text{Var}(n)/n^2 \approx \Delta \omega/\gamma \approx 1/n$ \cite{Fleckqto1966}, where $\Delta \omega$ is the laser linewidth and $\gamma$ the field relaxation time. Indeed, in the case of the superradiant laser, the linewidth is $\propto g^2/\kappa$ and, at large atom number and for $\kappa \gg \Gamma_R$, we have $\gamma_i \propto \kappa$. For standard lasers, the ultimate amplitude fluctuations arise because of fluctuations associated with spontaneous emission \cite{Fleck1966nln}, which introduces an amplitude drift that is damped by the cavity linewidth. The case of the superradiant laser using a beam architecture is different, since the stochastic process that introduces amplitude fluctuation is the outcoupling of atoms creating a drift of the collective atomic coherences. Far enough from the instability, the superradiant laser is still characterized by Poissonian fluctuations since $g^{(2)} (0) -1 =\frac{(\Delta n)^2 -<n>}{<n>^2} \propto  x \frac{4g^2}{\kappa \gamma_M}  -\frac{1}{<n>} \ll 1  $. However, when approaching the instability, intensity fluctuations are large and $g^2(0)$ significantly departs from 1.

Note that this discussion (which only takes into account the noise due to the out-coupling of atoms from the cavity) includes neither the quantum random noise associated with the introduction of the atoms in the cavity (as treated by TWA in sections of the present paper) nor that associated with the out-coupling of photons - which can be accounted for by a fluctuating Langevin force. Both would need to be included for a quantitative modeling.

\subsubsection{Laser linewidth}

The laser linewidth can be estimated by adding stochastic noises to the equations of the laser. This approach, based on the Langevin formalism, is shown in the Appendix of this paper. Our results for the cross-over  regime generalize the Shallow-Townes and superradiant limits. 

\begin{eqnarray}
    \Delta \nu &=& \frac{\kappa}{4 N_{\nu}} \frac{\Gamma_R^2}{\left( \Gamma_R+\frac{\kappa}{2} \right)^2} \left( 1 + \frac{N}{N_c} \right),
\end{eqnarray}
with the number of photons $N_{\nu} = \frac{g^2}{\kappa^2} \left( N-N_c \right) N_c$.
\subsection{Turbulent regime}

\subsubsection{Observation of a strange attractor} 

\begin{figure}
\centering
\includegraphics[width= 0.95\columnwidth]{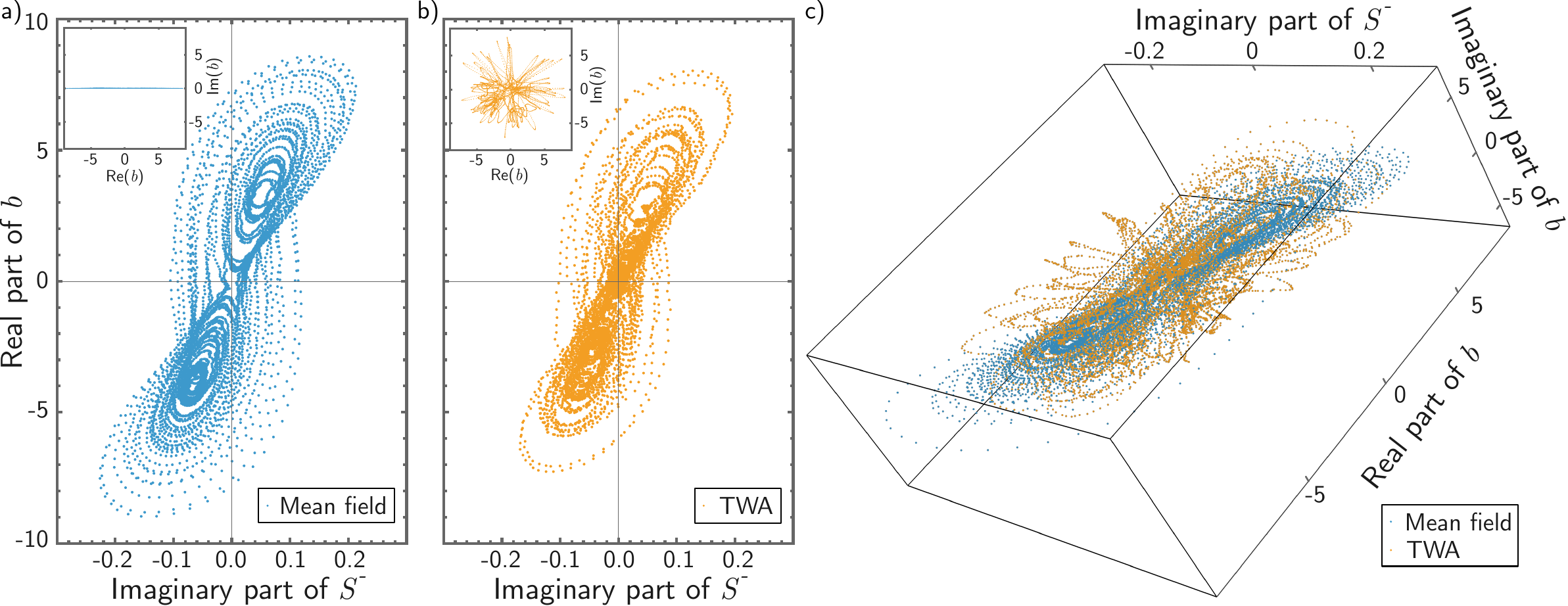}
\caption{\setlength{\baselineskip}{6pt} {\protect\scriptsize Trajectories during the superradiant laser dynamics, in the unstable regime. a, b) We represent the real part of the photon field $b$ versus the imaginary part of atomic coherences $S^-$. The laser parameters are $N = 400$, $g / 2 \pi = \sqrt{c} / 2$, $\kappa / 2 \pi =3$, $\Gamma_R / 2\pi = c$, with $c = 0.22$. The panel a) gives the mean-field approximation version (Eq.~\ref{averagedynamics})) and panel b) the version where we take into account atomic variable fluctuations using the TWA approximation in the Monte Carlo simulation using Eq.~\ref{elementarydynamics} (see main text). Insets show the same trajectories for the real and imaginary part of $b$, highlighting the effect of fluctuations on its phase. Panel c) shows the same two trajectories with the imaginary part of $b$ as a third axis.
}} 
\label{strangeattractor}
\end{figure}

To better understand the nature of the intensity instability that can occur when $\kappa > 4 \Gamma_R$, in the regime where the adiabatic approximation cannot be achieved, we show in Fig.~\ref{strangeattractor} a parametric plot representing the trajectories of the system (photon field \textit{vs} atomic coherences) as a function of time. We obtain these results just as in our previous Monte-Carlo simulations of Eqs.~\ref{elementarydynamics}, except that we plot the trajectory of a single realization. The panel Fig.~\ref{strangeattractor}a uses the mean-field approximation and an atom number high enough to be in the unstable regime. The figure suggests the existence of a strange attractor, where the dynamical state more or less regularly alternates between oscillating around one state and around a different one. Formally, the two points around which the system  oscillates irregularly are the two stationary points. Qualitatively, the oscillation rate around one unstable state is given by $\gamma_{osc}$ introduced above; the divergence rate from the unstable state is given by $\gamma_i$. When the oscillation is too large, the system jumps from one attractor to the other.  These observations confirm the analogy between laser dynamics and turbulence described above.

\subsubsection{Metrological implications} 
A natural question is that of the impact of turbulence onto the metrological quality of the superradiant laser when operated in the unstable regime. Interestingly, with the mean-field approximation the irregular behavior remains in a set plane, see Fig.~\ref{strangeattractor}c, which means that the irregular behavior does not introduce a phase drift. The respective phases of $b$ and $S^-$ simply jump by $\pm \pi$ when passing from an unstable state to the other. In this approximation, the turbulence only affects the amplitude of $b$ and not its phase, so the superradiant laser could remain a good active clock even in the presence of intensity fluctuations.

However, only some of our observations persist when we account for fluctuations of atomic variables, by using the Truncated Wigner Approximation (TWA). To simulate these fluctuations, we attribute random values $\pm 1/2$ to $s^x_M$ and $s^y_M$ for each new atom $M$ entering the cavity. As shown in the appendix, the phase diffusion of the laser is mostly set by the fluctuations of the atomic variables when the laser is operated sufficiently above lasing threshold. This is why we did not include the fluctuating terms associated with the photon leakage rate in these simulations. While this TWA approach does not drastically change the results of the simulations in terms of power and stability, it nevertheless introduces a drift of the phase of $\langle b \rangle$ into the complex plane. We show in Fig.~\ref{strangeattractor}b that the evolution of the photon field $\langle b \rangle$ remains strongly correlated with that of $S^- = \sum_j \langle s^-_j \rangle$. However, the apparent oscillations around two separate values cannot be clearly seen because of the phase drift, so that the system cannot be represented in a plane anymore. The representation of Fig.~\ref{strangeattractor}c highlights the presence of phase drifts while drawing a toroid-shaped kind of strange attractor.

A closer inspection of these trajectories reveals that the phase drifts are significant because of the turbulent character. When the amplitude of $b$ or $S^-$ approaches zero, the random jumps associated with the input and output of the atoms constitute a large modification of the phase of $b$ or $S^-$, respectively. Therefore, we expect the phase drift of the turbulent laser to be much faster than that of a superradiant laser in the stable regime, and we stress out that the turbulent regime must be avoided if the laser is to be used as an active optical clock. 

Some of these numerical observations can be supported by the phase equation Eq.~\ref{phaseequation} that is introduced in the appendix. While this equation can be used to predict the laser linewidth due to phase drift in the stable regime, it also signals that a given noise (due to the Langevin forces associated with photon out-coupling and atomic reloading) will affect $\rho \dot{\phi}$  (where $b = \rho e^{i \phi}$) so that the evolution of $\phi$ is stronger when $\rho$ approaches zero - which frequently happens in the fully developed turbulent regime.

\subsection{Oscillating regime}

In the chaotic regime described above, the system thus alternatively oscillates  around each of the two stationary points, and jumps from one attractor to the other in a rather unpredictable manner. As the atom number increases, the amplitude of the oscillations also increases, until a point where the jump between attractors occurs at each oscillation, so that in practice the oscillation encompasses both attractors. A self-pulsing regime is reached.

In this new regime, reached for very large values of $\Lambda = \frac{N}{N_c} - 1$, the system reaches another, rather regular, state in which the field remains real but periodically changes sign, at a regular rate of roughly $\sqrt{\Lambda} \Gamma_R$. The goal of this section is to provide a simplified description of this interestingly stable self-pulsing regime that arises significantly above the instability threshold.

Looking at the results of numerical simulations of Eq.~\ref{averagedynamicsrenorm} at large $\Lambda$, we posit for $\Sigma^z(t)$ the Ansatz $\Sigma^z(t) = \Lambda \frac{\Gamma_R}{\kappa} (1- \beta(t)^2) + 2 \Lambda \frac{\Gamma_R^2}{\kappa^2}$ and look for an approximate analytical solution for $\beta(t)$. This choice of Ansatz will be shown to be self-consistent at the end of this section, by verifying that the analytical expression $\beta(t)$ approximates well the numerical solution of Eq.~\ref{averagedynamicsrenorm}. We also numerically find a roughly periodic behavior in time, which will have an impact on the approximations made in the following.

We point out that, if the dynamics of the system corresponds to an oscillation of all parameters at $\gamma_{osc} \approx \sqrt{\Lambda} \Gamma_R$ around the stationary points (see stability analysis above), we generically have, for large $\Lambda$, in terms of orders of magnitude, $\langle {\Sigma^-}^2\rangle \propto \frac{4 \Gamma_R^2 \Lambda}{\kappa^2} \langle \beta^2\rangle$ and $\langle (\Sigma^z \beta)^2\rangle \propto \Lambda \langle {\Sigma^-}^2\rangle$. Therefore, at large $\Lambda$, in terms of orders of magnitude, $\Sigma^z \beta \gg \Sigma^-$ and $\Sigma^- \approx 2 \beta'/\kappa$. We thus have:

\begin{eqnarray}
    \beta''(t) &=& -\frac{\kappa}{2} (\beta'-{\Sigma^-}') =  -\frac{\kappa}{2} \beta' +\frac{\kappa \Gamma_R}{2} (\Sigma^z \beta - \Sigma^-) \nonumber \\*
    &\approx&   -\frac{\kappa}{2} \beta'(t) +  \Lambda \frac{\Gamma_R^2}{2} (1- \beta^2)\beta  + \Lambda \frac{\Gamma_R^3}{\kappa} \beta.
    \label{approxE}
\end{eqnarray} 
If $\beta$ oscillates at a frequency $\sqrt{\Lambda} \Gamma_R$, the order of magnitude of $\beta'$ is $\propto \sqrt{\Lambda} \Gamma_R \beta$. Therefore the first term of the right-hand-side of Eq.~\ref{approxE} can be neglected when $\sqrt{\Lambda} \frac{\Gamma_R}{\kappa} \gg 1 $, in which case: 

\begin{equation}
   \beta''(t) \approx \Lambda \frac{\Gamma_R^2}{2} \left( (1- \beta^2)\beta  + 2\frac{\Gamma_R}{\kappa} \beta\right).
   \label{approxE2}
\end{equation}

This expression shows that the system is governed by a single frequency $ \sqrt{\frac{\Lambda}{2}} \Gamma_R$ when $\Gamma_R \ll \kappa$ (in which case the last term of the equation can be neglected). An oscillating regime is thus reached when the system is sufficiently in the bad-cavity regime and for a large atom number. The criterion $\sqrt{\Lambda} \frac{\Gamma_R}{\kappa} \gg 1 $ is a good indicator of the crossing from chaotic to oscillating as $N$ (and $\Lambda$) increases. The equation Eq.~\ref{approxE2} is that of an inverted pendulum (non-damped Duffing equation) and has for solution (assuming $\beta(0)=0$):
\begin{equation}
    \beta(t)=-i \sqrt{-A+\sqrt{A^2+2 c_1}} \text{sn}\left(\sqrt{-\frac{\Lambda\Gamma_R^2}{4}t^2 \left(A+\sqrt{A^2+2
   c_1}\right)}|-\frac{A^2-\sqrt{A^2+2 c_1} A+c_1}{c_1}\right)
   \label{beta}
\end{equation}
with $A=1+\frac{2 \Gamma_R}{\kappa}$; $\text{sn}$ is the Jacobi elliptic function. $c_1$ is a constant that sets $\beta'(0)=\sqrt{c_1}$, and the shape of $\beta(t)$. To find the value of $c_1$, we compute $<\beta^2>$ and find

\begin{equation}
    <\beta^2> = \frac{ \left(A+\sqrt{A^2+2 c_1}\right) \left(i \mathcal{E}\left(2 i B|-\frac{A^2-\sqrt{A^2+2 c_1} A+c_1}{c_1}\right)+2 B\right)}{2  B}
    \label{E2}
\end{equation}
where $B=\Re\left(K\left(\frac{2}{\frac{A}{\sqrt{A^2+2 c_1}}+1}\right)\right)$,
$\mathcal{E}$ stands for the Jacobi Epsilon function, $K$ is the complete elliptic integral of the first kind, and $\Re$ the real part.

We also have, from the last equation of Eq.~\ref{LHmodel}, $\Lambda+1 -<\Sigma^z>-\Lambda <\beta \Sigma^->=0$, which implies, for $\Lambda \gg 1$ (and using the first equation in Eq.~\ref{LHmodel}, as well as the periodicity of $\beta$, so that $<\frac{d\beta}{dt}> =0$):
\begin{equation}
<\beta^2> \approx \frac{ 1 - \frac{\Gamma_R}{\kappa} - 2  \frac{\Gamma_R^2}{\kappa^2}}{1 -  \frac{\Gamma_R}{\kappa}}.
\label{E2_bis}
\end{equation}

Equating Eq.~\ref{E2} and Eq.~\ref{E2_bis}, we find $c_1=f\left(\frac{\Gamma_R}{\kappa}\right)$. Although we were unable to provide an analytical solution for $f$, we find (using expansion up to the second order in $\frac{\Gamma_R}{\kappa}$): $f(x) \approx  0.68 -3.04 x + 1.37 x^2$.

\begin{figure}
\centering
\includegraphics[width= 0.95\columnwidth]{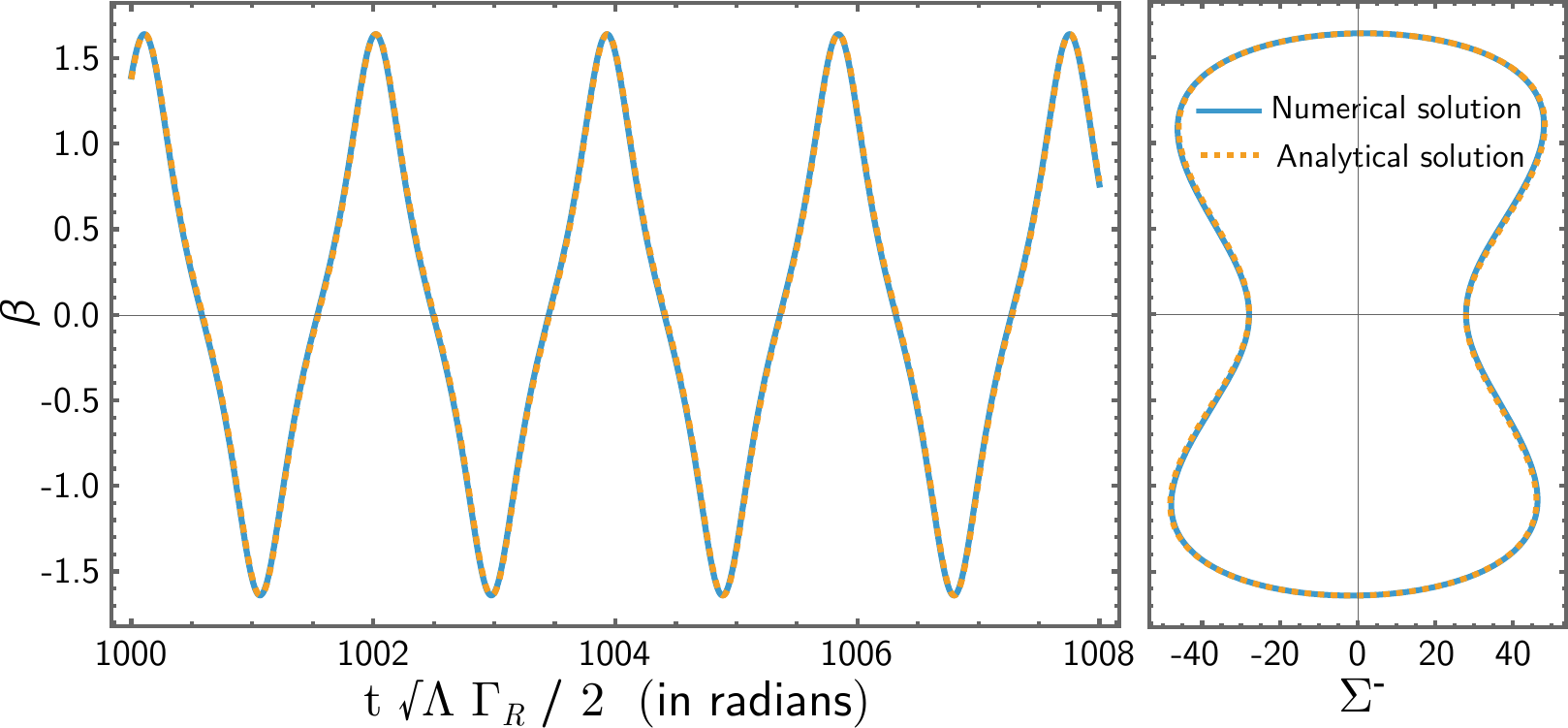}
\caption{\setlength{\baselineskip}{6pt} {\protect\scriptsize Comparison between numerical MF simulations (blue solid line) of Eq.~\ref{averagedynamicsrenorm} and the analytical model (Eq.~\ref{beta}, dashed orange curve) at large atom numbers ($\Lambda = 10^5$), for $\kappa / 2 \pi = 2$, $\Gamma_R / 2 \pi = 0.2$. The left panel shows $\beta$ as a function of time. The (irrelevant) initial phase of the oscillation for the numerical simulation has been shifted to provide the best agreement between both curves. The right panel shows that the system follows a regular, non-sinusoidal oscillation enclosing both steady-state solutions at $(1,1)$ and $(-1,-1)$. Note that the shape of the oscillation depends on its amplitude, and that both shape and amplitude are controlled by the value of $c_1$ (see text).}} 
\label{beyondturbulence}
\end{figure}

As an example, when $\kappa / 2\pi=2$ and $\gamma_R / 2\pi=0.2$ (and $\Lambda=10^5)$), using $c_1=f(0.1)=0.39$, we find nearly perfect agreement between direct numerical solutions of Eq.~\ref{averagedynamicsrenorm} and the analytical model, as shown in Fig.~\ref{beyondturbulence}.

This analysis confirms that a rather regular and well-defined oscillation is obtained at very large atom numbers, even in the regime where turbulent behavior is seen at intermediate atom numbers. However, we still expect that the phase drift of the laser (and therefore its linewidth) will be considerably affected in this regime.

\section{Conclusions}

This paper investigates the stability of the superradiant laser based on an architecture in which the atomic medium is constantly refilled. We have used Monte-Carlo simulations as well as a simplified model that show that the superradiant laser can become unstable owing to a dynamical instability. In the mean-field approximation, the instability formally maps to the Hakken-Lorenz model for turbulence. However, taking into account fluctuations by means of a Langevin fluctuating force significantly enriches the dynamical behaviors. The regime in which such instabilities are expected to occur is relevant for some experiments being built. Our work thus helps in defining which experimental parameters are best for metrological applications, especially since we find that the laser becomes noisy significantly before it crosses the regime of instability. 
Indeed, at worst $N_c^*/N_c$ is on the order of 15, which is reached for $\kappa/\Gamma_R \simeq 11$, leaving little room in the variable $N$ for stable superradiant lasing. Nevertheless, optimizing the experimental design for metrological performance, and in particular to minimize the cavity pulling coefficient, leads to $\kappa \gg \Gamma_R$ and increases the margin for stable superradiant lasing. For example, in the theoretical proposal \cite{Liu2020RuggedSuperRadiantOven}, two working condition examples for an atomic beam superradiant laser are considered, for which $N_c^*/N_c = 500$ and $N/N_c^* \simeq 0.1$. This motivates further quantitative study of intensity fluctuations, precursors to the instability into the stable regime, and of how detrimental they are to metrological applications.

Furthermore, our work  paves the way for future investigations of the unstable regime. Among the interesting questions ahead: can one calculate the linewidth of the superradiant laser when operated in the chaotic regime? What would be the $g_2$ of the chaotic laser?  Is the chaotic laser still useful from the metrological perspective, and how far from the instability must one stay to preserve a narrow linewidth? It would also be interesting to use the second order cumulant approach in order to include correlations in the system and estimate the linewidth \cite{Dubey2024PhDThesis}, and to assess whether the quantum nature of the system modifies the scenario of  turbulence that has been discussed throughout this paper.


\section{Appendix on the laser linewidth}

To derive the laser linewidth, we complement the laser equations Eqs.~\ref{averagedynamics} with fluctuating Langevin forces \cite{Sargent1977, Benkertrop1990} that are meant to account for the drift of the observables due to the stochastic out-coupling of photons, and in-and-out-coupling of atoms:

\begin{eqnarray}
    i  \frac{db}{dt} &=& g S^-  -\frac{i}{2}  \kappa  b + F_b(t) \nonumber \\
   i  \frac{dS^-}{dt} &=&  -2 g  b S^z - i \Gamma_R S^- + F_-(t) \nonumber \\
 i  \frac{dS^z}{dt} &=&  g b S^+ - g  b^{\dag}  S^- - i \Gamma_R S^z+ \frac{i}{2} \Gamma + F_z(t)
 \label{averagedynamics2}
 \end{eqnarray}

\subsection{Phase equation}

\subsubsection{Derivation of the phase equation} 

We apply time-derivation on the first line of Eqs.~\ref{averagedynamics2}, and use its first two lines to express $b$ only as a function of the fluctuating forces and $S^z$. We write $b= \rho  e^{i \phi}$. We then take the real part ($\Re$) of the resulting equation and find:
\begin{equation}
    - 2 \dot{\rho} \dot{\phi} - \rho \ddot{\phi} = \left( \frac{\kappa}{2} + \Gamma_R\right) \dot{\phi} \rho +\Re \left( \left( \dot{F_b} + \Gamma_R F_b +\frac{g F_-}{i} \right) e^{-i \phi}\right) ,
    \label{phaseequation}
\end{equation}
which, interestingly, does not depend on $S_z$.

\subsubsection{Approximations and simplified phase equation} 

We now make two assumptions. 

First, we assume $\frac{\dot{\phi}}{\phi} \ll \frac{\kappa}{2}+\Gamma_R$, implying that the laser linewidth is sufficiently small, which will be verified to be self-consistent at the end of the calculation. 

Second, we assume $\frac{\dot{\rho}}{\rho} \ll \frac{\kappa}{2}+\Gamma_R $. This assumption basically means that the laser intensity is supposed to be sufficiently stable and that the characteristic time evolution of the density (set by the eigenvalues of $M$) are small compared to $\frac{\kappa}{2}+\Gamma_R$. 

Thanks to these assumptions, and provided their domain of validity is respected, we find the following simplified equation, which we will use to derive the laser linewidth.

\begin{equation}
     \dot{\phi} \rho = -\frac{1}{\left( \frac{\kappa}{2} + \Gamma_R\right)}\Re \left( \left( \dot{F_b} + \Gamma_R F_b -ig F_- \right) e^{-i \phi}\right)
     \label{simplifiedphase}
\end{equation}

The assumptions leading to Eq.~\ref{simplifiedphase} are not straightforward: 
\begin{itemize}
    \item We first examine the case of the superradiant regime, and the effect associated with the atomic variable fluctuations. In the case where the adiabatic elimination of the field is ensured, $b$ is anchored to the evolution of $S^-$. $S^-$ includes fluctuations due to the in-coupling and out-coupling of atoms, at a rate $\Gamma$. There are therefore fluctuations $\frac{\delta b}{\delta t} \approx \frac{g}{\kappa} \Gamma$, so that $\frac{\dot{\rho}}{\rho} \approx  \frac{4 \Gamma_R N}{\sqrt{N_c(N-N_c)} }$. Therefore, the approximation $\frac{\dot{\rho}}{\rho} \ll \frac{\kappa}{2}+\Gamma_R $ is not verified at a large number of atoms, or close to the threshold.
    \item We now further examine the case of the superradiant regime and the effect associated with the photonic variable fluctuations, and will argue that the approximations made above are then incompatible with the adiabatic elimination of the field. The adiabatic approximation of $b$ relies on a separation of timescales. $b$ needs to remain anchored to the (slow) evolutions of $S^-$, so that $g S^-  -\frac{i}{2}  \kappa  \bar{b} = 0$ with $\bar{b}$ the average value of $b$. This equality must be fulfilled for timescales $\frac{1}{\kappa} <\tau< \frac{1}{N\Gamma_R}$. The left-hand-side inequality represents the damping time for the photon variable; the right-hand-side inequality is because at each insertion time $\frac{1}{\Gamma_R}$ the system is perturbed by the addition and loss of a new atom: using a coarse-graining over a timescale $\tau>\frac{1}{N\Gamma_R}$ would not ensure a proper estimate of $\frac{db}{dt}$ since the dynamical evolution of the atomic degrees of freedom is then non negligible. Within this time window, the fluctuations are set by $\frac{db}{dt} = F_b(t)$. We now consider the typical coarse-graining time $\tau$ over which fluctuations of the field need to be considered so that the approximation $\frac{db}{bdt} \ll \kappa$ holds. We have $\frac{\delta b}{dt} = F_b(t) \approx \sqrt{\frac{\kappa}{\tau}} $ so that $\frac{db}{bdt} \approx \sqrt{\frac{\kappa}{\tau}} \frac{\kappa}{g S^-} \approx \sqrt{\frac{1}{\tau}} \frac{\kappa }{ \sqrt{N \Gamma_R}} $. Therefore, $\frac{\delta b}{\delta dt} \ll \kappa$ implies a coarse-graining time $\tau> \frac{1}{ \Gamma_R N}$, which is incompatible with the allowed time window   $\frac{1}{\kappa} <\tau< \frac{1}{N\Gamma_R}$. As a consequence, Eq.~\ref{simplifiedphase} cannot be used in the superradiant regime where adiabatic approximation is typically performed.
    \item In the dynamically unstable regime, at a large number of atoms where the pulsed regime is reached, $\frac{\delta \rho}{\delta t} \approx \rho \Gamma_R \sqrt{\frac{N}{N_c}}$. Again, the approximation is not verified.
    \item On the other hand, in the stable regime, where intensity fluctuations are small and when the adiabatic elimination is not performed the intensity fluctuations are typically $\frac{\delta \rho}{\rho} \approx \frac{g^2}{\kappa \gamma_M}$ and occur over timescale $\gamma_M^{-1}$. Therefore, the assumption is verified as along as $\frac{g^2}{\kappa} \ll  \frac{\kappa}{2}+\Gamma_R $.
\end{itemize}.

\subsection{Amplitude of the fluctuating forces}

The functions $F_{b,-}(t)$ are discussed in many references \cite{Haken1986,Sargent1977,Benkertrop1990,Tieri2017CrossoverLasingSuperradiance}. Here, we first give a practical heuristic argument to define their correct value for a given numerical simulation. In a practical simulation, any random function must be defined within a given bandwidth. We therefore assume that  $F_{b,-,z}(t)$ consist of random values (described by a Gaussian white noise of a given amplitude  $A_{b,-,z}(t)$) that are modified at each time increment $\delta t$. The amplitude $A$ is chosen to reproduce the expected drift of the variable $b,S^-,S^z$.

To correctly describe the fluctuation of the variables, one has to understand the various stochastic mechanisms and characterize them by proper stochastic functions for the different observables, that can be correlated from one observable to the other. It turns out that the beam architecture for the superradiant laser - neglecting spontaneous emission, the decoherence, and without repumping - is rather simple to describe. 

There are three main stochastic processes. The first one is the photon out-coupling, and is uncorrelated to the stochastic processes affecting the atomic variables. The second is the insertion of atoms. This insertion of atoms is done with atoms purely in $S_z=1/2$. Therefore, it introduces no fluctuation for this variable. However, it introduces fluctuations for $S_x$ and $S_y$. Those can be properly accounted for by a TWA approach, which corresponds to an uncorrelated diffusion rate of $\Gamma /4$ in the $x$ and $y$ directions only. Finally, the out-coupling of atoms also introduces fluctuations, and is the most subtle to tackle. An impact is on $S_z$ and need not be described here because fluctuations in $S_z$ do not affect the phase, following Eq.~\ref{simplifiedphase}. Another impact is on the atomic dipole. The key point is that, in the steady state, the dipole is perpendicular to the field $b = -i \frac{2g}{\kappa} S^-$. In Eq.~\ref{simplifiedphase}, it appears that the important fluctuations to take into account are fluctuations of the dipoles perpendicular to the main dipole, $i.e$ parallel to the field. Let us call $Y$ this direction at a time $t$. $\langle S_Y \rangle=0$, therefore, what counts is the variance in the $Y$ direction, which is exactly $1/4$ per atom, at a rate $\Gamma$.

Therefore, the total effect of in-coupling and out-coupling on the system can be described by an irrelevant fluctuation of $S_z$, and an uncorrelated increase of the variance $\text{Var}(S^-)$ with a rate $ = 2 \times \frac{\Gamma}{4}$, leading to 

\begin{equation}
    F_-(t) =  \sqrt{ \frac{\Gamma}{2 \delta t}} a_- \left( \lceil \frac{t}{\delta t} \rceil  \right)
    \label{diffusinom}
\end{equation}
where $ \lceil \frac{t}{\delta t} \rceil$ represents the integer part, and the function $a_- \left( \lceil \frac{t}{\delta t} \rceil  \right)$ generates random values of Gaussian statistics, of variance 1, that are modified at each integer value of $ \frac{t}{\delta t}$.

More formally, we  point out the mapping between the continuous beam and the repumped superradiant laser. The mean equations are identical provided that we set $\Gamma_1/2=\Gamma_R$ and $\Gamma_2=\Gamma_R$. This implies that $R+\nu_s =\Gamma_R$ and the additional decoherence is $\nu=\Gamma_R$. This is  what the TWA approximation states: at the end of each insertion time $1/\Gamma$, an atom disappears, and an atom appears in the upper state (which exactly mimics repumping), and in addition a random noise $\pm 1/2$ (TWA) is added to $S_x$ and $S_y$ \cite{Schachenmayer2015QuSpinDyna}. 

Using this exact mapping, we also refer to \cite{Sargent1977}, and write $G=\frac{1}{\Gamma_R+\frac{\kappa}{2}} \left(\Gamma_R F_b - i g F_- \right)$. We therefore have $\frac{d \phi}{dt} = \frac{1}{\rho} \Re (G e^{-i \phi)})$ (discarding the influence of $\dot{F_b}$ in Eq.~\ref{simplifiedphase} which does not introduce drift, see below) and the phase diffusion coefficient is shown to be: $2 D_{\phi} = \frac{1}{4 \rho^2} \langle G^+ G+G G^+ \rangle \delta t$ \cite{Sargent1977}. We point out that $\langle F^+(t) F^-(t') \rangle  + \langle F^-(t) F^+(t') \rangle =2 \langle F^x(t) F^x(t') \rangle +2 \langle F^y(t) F^y(t') \rangle$. By symmetry, $\langle F^x(t) F^x(t') \rangle=\langle F^y(t) F^y(t') \rangle =2 D_{xx} \delta (t-t') = 2 D_{yy}  \delta (t-t') $. Given the values of $D_{-+}$ and $D_{+-}$ in \cite{Tieri2017CrossoverLasingSuperradiance}, we have $2 D_{xx} = 2 D_{yy} = \Gamma_1 N /4 =N \Gamma_R/2$.  The drift associated with the atomic noise is thus set by  $2D_a=\frac{g^2}{\rho^2 \left(\Gamma_R+\frac{\kappa}{2}\right)^2} \frac{\Gamma_R N}{2}$. This formally justifies the intuitively introduced Eq.~\ref{diffusinom} for the fluctuating variable $F_-(t)$, that does introduce the correct phase diffusion.

In addition, $\langle F_b^+(t) F_b^-(t') \rangle +\langle F_b^-(t) F_b^+(t') \rangle =\langle F_b^-(t) F_b^+(t') \rangle $ at zero temperature \cite{Haken1986, Sargent1977,Benkertrop1990}. Therefore the phase drift associated with the leaking of photons from the cavity is $2D_{\nu}=\frac{\Gamma_R^2}{\rho^2 \left(\Gamma_R+\frac{\kappa}{2}\right)^2} \frac{\kappa}{4}$. In practice, this is described in a numerical simulation by a noise given by 
\begin{equation}
    F_b =  \sqrt{\frac{\kappa}{4 \delta t}} a_b \left( \lceil \frac{t}{\delta t} \rceil  \right) ,
\end{equation}
where  the function $a_b \left( \lceil \frac{t}{\delta t} \rceil  \right)$ generates random values of Gaussian statistics, of variance 1, that are modified at each integer value of $ \frac{t}{\delta t}$.

\subsection{Laser linewidth in the stable regime}

In Eq.~\ref{simplifiedphase}, we discard the influence of $\dot{F_b}$. Indeed, when time-integrating Eq.~\ref{simplifiedphase}, the effect of $\dot{F_b}$ is bounded and does not introduce drift. We now evaluate the influence of $F_b$ and $F_-$ separately. We assume that $\rho$ is constant.

\subsubsection{Influence of the photon fluctuation}

Over a small duration $\delta t$, there is a typical phase increment $\delta \phi \approx \delta t \frac{\Gamma_R \sqrt{\kappa}}{\rho \left( \Gamma_R+\frac{\kappa}{2} \right) \sqrt{4 \delta t}}$, which is an individual step of a random walk repeated a number of $\frac{t}{\delta t}$ times after a duration $t$. Therefore $\text{Var}(\phi) \approx \frac{\Gamma_R^2 \kappa}{4 \rho^2 \left( \Gamma_R+\frac{\kappa}{2} \right)^2 } t $ and the laser linewidth is
\begin{equation}
    \Delta \nu_1 \approx\frac{\kappa}{4 \rho^2}\frac{\Gamma_R^2}{\left( \Gamma_R+\frac{\kappa}{2} \right)^2 }.
\end{equation}
We recover the Shallow-Townes limit, with the correction due to narrow gain when in the bad-cavity regime $\kappa \gg \Gamma_R$.

\subsubsection{Influence of the atomic coherence fluctuation}

Over a small duration $\delta t$, there is a typical phase increment $\delta \phi \approx \delta t \frac{g \sqrt{\Gamma}}{\sqrt{2} \rho \left( \Gamma_R+\frac{\kappa}{2} \right) \sqrt{\delta t}}$, which is an individual step of a random walk repeated a number of $\frac{t}{\delta t}$ times after a duration $t$. Therefore $\text{Var}(\phi) \approx \frac{g^2 \Gamma}{2\rho^2 \left( \Gamma_R+\frac{\kappa}{2} \right)^2 } t $ and the laser linewidth is
\begin{equation}
    \Delta \nu_2 \approx\frac{g^2}{\rho^2}\frac{\Gamma}{2\left( \Gamma_R+\frac{\kappa}{2} \right)^2 }.
\end{equation}
At equilibrium when $g S^-=-\frac{i}{2} \kappa b$, and using Eq.~\ref{solstat} for $\lvert S^-\rvert^2$, we find 
\begin{equation}
    \Delta \nu_2 \approx\frac{g^2}{ \kappa}\frac{\kappa^2}{\left( \Gamma_R+\frac{\kappa}{2} \right)^2 } \frac{N}{N-N_c},
\end{equation}
which corresponds to the ultimate laser linewidth of the superradiant laser.

\subsubsection{Discussion}

\begin{itemize}
    \item  The condition $\Delta \nu_{(1,2)} \ll \left( \frac{\kappa}{2}+\Gamma_R \right)$ is easily satisfied, which ensures the previously assumed approximation $\dot{\phi} \ll \frac{\kappa}{2}+\Gamma_R$.

    \item We have $\frac{\Delta \nu_2}{\Delta \nu_1}=  \frac{N}{N_c}$. Therefore, we expect the laser linewidth is mostly set by the photon out-coupling close to threshold, and mostly set by the atomic fluctuations at large atom number.  

    \item Our result can be compared to the Shallow-Townes limit reached in the good-cavity regime, where adiabatic elimination of the atomic coherences is performed, in which case $\Delta \nu_{ST} =\frac{\kappa}{4  \rho^2} \left( 1 + \frac{N}{2 S_z}\right) $ \cite{Sargent1977}, which is equal to $\Delta \nu=\Delta \nu_1 + \Delta \nu_2$ when $\Gamma_R \gg \kappa$.

    \item On the other hand, in the bad-cavity regime, the field $b$ is bound to the dipole $S^-$. We have $\frac{dS^-}{dt} = \frac{4g^2}{\kappa} S^- S^z - \Gamma_R S^- +F_-$. Writing $S^-=\vert S^- \rvert e^{i \theta}$ and taking the imaginary part of the equation, we have $ \dot{\theta} \vert S^- \rvert = \Im \left( F_- e^{-i \theta} \right)$, from which we deduce that the linewidth of the laser is $\frac{2 g^2}{\kappa} \approx_{\kappa \gg \Gamma_R} \Delta \nu_2$.

    \item In the crossover regime, the linewidth is given by the usual Shallow-Townes formula, except that it is modified by the factor $\frac{\Gamma_R^2}{\left( \Gamma_R+\frac{\kappa}{2} \right)^2}$. This correction can be attributed to the dispersion associated with the medium of finite width $\Gamma_R$, which slows the light \cite{Exter1995LinewidthBadCavityLaser}. The superradiant limit, for which the linewidth is set by the Purcell rate, can be seen as the large $\kappa$ and large $N$ limit of $\Delta \nu$. 
    
\end{itemize}

\subsection{Laser linewidth in the presence of intensity fluctuations}

It can be seen  from Eqs.~\ref{phaseequation} and \ref{simplifiedphase} that, when the laser intensity is fluctuating, the discussion is more complicated. However, it is clear that when $\rho$ dynamically approaches 0, the values of $\dot{\phi}$ will get very large, and the impact of the fluctuating forces exacerbated. Therefore a natural outcome of this analysis is that the unstable regime must be avoided if the laser is to be used as a frequency reference.

\section*{Acknowledgements}
We thank Yannis Pargoire and Zeyuan Zhang for stimulating discussions. 

\paragraph{Materials}
All the code, simulation results, and figures used for this work can be found at the repository: https://src.koda.cnrs.fr/lpl/public/gqm-public/paper-instability-cw-sr-laser

\paragraph{Funding information}
We acknowledge support from Agence Nationale de la Recherche (project CONSULA, ANR-21-CE47-0006-02), Labex FIRST-TF (projects SURECO and COCO), and Ile de France Region in the framework of DIM SIRTEQ (project FSTOL) and DIM Quantip (project STACS). This work is supported by France 2030 QuanTEdu program and the European Union NextGenerationEU program, under ANR n° 22-CMAS-0001.


\end{document}